\documentclass[10pt,journal,compsoc]{IEEEtran}
\usepackage[utf8]{inputenc}
\pdfoutput=1

\usepackage[hyphens]{url}
\usepackage[hidelinks]{hyperref}

\usepackage{booktabs}
\usepackage{graphicx}

\ifCLASSOPTIONcompsoc
  \usepackage[nocompress]{cite}
\else
  \usepackage{cite}
\fi

\author{Lei~Zhang,
        Andriy~Miranskyy,
        and~Walid~Rjaibi
\IEEEcompsocitemizethanks{\IEEEcompsocthanksitem L.~Zhang and A.~Miranskyy are with the Department
of Computer Science, Ryerson University, Toronto, Canada.\protect\\
E-mails: leizhang@ryerson.ca and avm@ryerson.ca
\IEEEcompsocthanksitem W.~Rjaibi is with IBM Canada Lab, Toronto, Canada.\protect\\
E-mail: wrjaibi@ca.ibm.com}
}

\title{Quantum Advantage and Y2K Bug: Comparison}

\begin{document}

\IEEEtitleabstractindextext{
\begin{abstract}
Quantum Computers (QCs), once they mature, will be able to solve some problems faster than Classic Computers. This phenomenon is called ``quantum advantage'' (which is often used interchangeably with a stronger term ``quantum supremacy''). 

Quantum advantage will help us to speed up computations in many areas, from artificial intelligence to medicine. However, QC power can also be leveraged to break modern cryptographic algorithms, which pervade modern software: use cases range from encryption of Internet traffic, to encryption of disks, to signing blockchain ledgers. 

While the exact date when QCs will evolve to reach quantum advantage is unknown, the consensus is that this future is near. Thus, in order to maintain crypto agility of the software, one needs to start preparing for the era of quantum advantage proactively (before the software and associated data are compromised). 

In this paper, we recap the effect of quantum advantage on the existing and new software systems, as well as the data that we currently store. We also highlight similarities and differences between the security challenges brought by QCs and the challenges that software engineers faced twenty years ago while fixing widespread Y2K bug. Technically, the Y2K bug and the quantum advantage problems are different: the former was caused by timing-related problems, while the latter is caused by a cryptographic algorithm being non-quantum-resistant. However, conceptually, the problems are similar: we know what the root cause is, the fix (strategically) is straightforward, yet the implementation of the fix is challenging.

To address the quantum advantage challenge, we create a seven-step roadmap, deemed 7E. It is inspired by the lessons-learnt from the Y2K era amalgamated with modern knowledge. The roadmap gives developers a structured way to start preparing for the quantum advantage era, helping them to start planning for the creation of new as well as the evolution of the existent software.

\end{abstract}
}

\maketitle

\IEEEraisesectionheading{\section{Introduction}\label{sec:introduction}}
\IEEEPARstart{T}{he} field of quantum computing is still young, but it has been evolving rapidly during the last decade. With the vast increase in computing power, Quantum Computers (QCs) promise to revolutionize many fields, including artificial intelligence, medicine, and space exploration. However, they can also be abused to break key encryption algorithms the Internet depends upon today for ensuring the safety and privacy of digital information. Thus, we position that the software engineering community should start thinking about the impact of quantum computing on cybersecurity and the best practices to address these concerns. Let us look at the evolution of QCs.

\subsection{Evolution}
In 1982, Feynman introduced the idea of quantum computing~\cite{feynman1982simulating}; Shor proposed the first practically relevant algorithm (for breaking encryption protocols based on integer factorization and discrete logarithm) that can be efficiently computed on a QC in 1994~\cite{shor1997}.

It took a while to implement an actual QC. A partnership between academia and IBM created the first working 2-qubit QC in 1998~\cite{Chuang1998}, but it took the company 18 years to make a 5-qubit QC accessible to the public in 2016~\cite{ibm2016}. 

At present, a few QCs are commercially available. D-Wave started selling adiabatic QC in 2011 (although the debate about adiabatic QC being a ``true'' QC is ongoing\footnote{A hybrid of adiabatic and gate-based QC is promising~\cite{barends_digitized_2016}, but no commercial implementation is available.}~\cite{albash2017}) with the current offerings having $>$~2000 qubits. IBM gave access to  20- and 50-qubit gate-based superconducting QCs to academic and industrial partners to explore practical applications in 2017~\cite{ibm2017}. 

For non-commercial use, IBM offers 5- and 14-qubit QCs via IBM Q Experience online platform based on IBM Cloud (along with local- and Cloud-based simulators)~\cite{ibm_quantum}. Microsoft provides access to a simulator of a topological QC via Microsoft Quantum Development Kit~\cite{ms_quantum} (and is planning to give access to an actual QC in the future). Google built 72-qubit gate-based superconducting QC in 2018~\cite{google2018}, but it is not publicly accessible at the time of writing. 

\subsection{Quantum advantage: timeline}
It is said that in the future, a QC can solve some problems much faster than a Classical Computer (CC), which is called quantum advantage~\cite{feynman1982simulating} (often used interchangeably with the term quantum supremacy, which denotes an ability of QCs to solve problems that CCs cannot). This is because QC compute power is growing faster than CC one. 

The exact growth rate of QCs is under debate. Norishige Morimoto, the director of IBM research in Tokyo and global vice president at IBM, claims that quantum advantage will be achieved between years 2022--2024~\cite{ibm2022}. Michelle Mosca, a co-founder of the Institute for Quantum Computing and chief executive of evolutionQ, claims that QC may outperform CC in certain tasks after 2026~\cite{mosca2018cybersecurity}. 

One of the measurements of the QC performance\footnote{The performance of the QCs, based on different architectures, cannot be compared merely based on the number of qubits that each QC has. This is akin to difficulty in comparing the performance of the central processing unit (CPU) of a CC based only on the number of CPU cores and the cores' frequency.}, introduced by IBM, is the Quantum Volume. To achieve quantum advantage within the next decade, IBM stated that they ``need to at least double the Quantum Volume of our quantum computing systems every year.~\cite{chow2020}'' In January of 2020, Chow and Gambetta confirmed that IBM is on track to reach this goal with a new 28-qubit QC demonstrating the Quantum Volume of 32~\cite{chow2020}. 

\subsection{Quantum advantage: cyberthreat}
The power of quantum computing will threaten modern cybersecurity platforms by speeding up 1) factorization of integers, solving the discrete logarithm problem, and the elliptic-curve discrete logarithm problem (using Shor’s algorithm~\cite{shor1997}); as well as 2) the search in a set  (with the help of Grover’s algorithm~\cite{grover1996fast}). Both tasks are foundational for modern encryption algorithms. 

\begin{figure}
	\centering
	\includegraphics[width=0.46\textwidth]{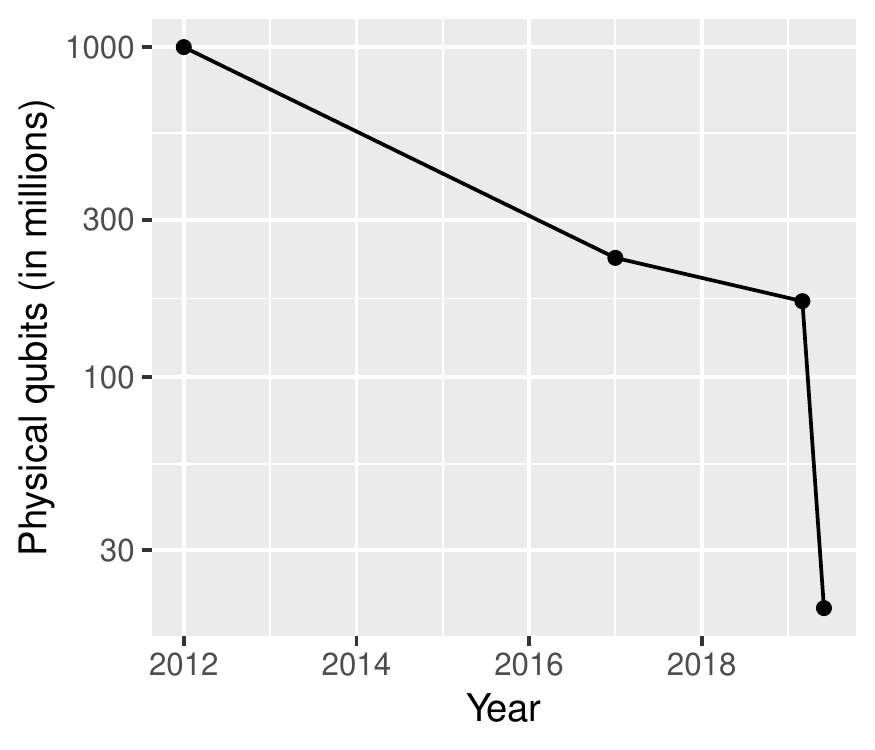}
	\caption{Number of qubits (in millions) needed to break RSA 2048 based on the literature review in~\cite[Table II]{gidney2019factor}. The x-axis denotes a year when an article was published.
	}
	\label{fig:20milli}
\end{figure}

While the existing QCs are not ready to use Shor’s algorithm for production problems, a lot of brilliant people (both on the hardware and the algorithm sides) are racing to get us there. For example, Gidney et al.~\cite{gidney2019factor} combine a set of clever tricks and techniques to implement Shor’s algorithm using modern QC architecture. They show that one needs 20M qubits to break the 2048-bit RSA key in less than a day. They compare their solution against an existing one by Fowler et al.~\cite{Fowler2012Surface}, which was introduced in 2012. This baseline approach required one billion qubits. Thus, in seven years, algorithms designers reached a remarkable 165x ~\cite[Table II]{gidney2019factor} improvement in the algorithm implementation. This dynamic is represented graphically in Figure~\ref{fig:20milli}. The 20M qubits is not an optimal boundary, which implies that in the future, improvement schemes may be derived, further reducing the hardware requirements.

On the hardware side, the electrical engineers and physicists will continue working on improving coherence, noise reduction, and error correction solutions, making QCs more reliable and robust. In theory~\cite{qubits4099}, a QC with 4099 perfectly stable qubits can break RSA 2048 in 10 seconds. While we doubt that such stability can be achieved soon, the hardware systems are improving. Moreover, new contenders in this field (such as photonic quantum computer developed by Xanadu~\cite{xanadu}) may introduce new and efficient computing architectures that will enable algorithm designers. 

\subsection{Quantum advantage: call to action}\label{sec:cta}

While the date when quantum advantage will be achieved may be several years away, we need to start protecting ourselves against this eventuality right now. This is because malicious entities can harvest sensitive data communications now and~--- when a powerful quantum computer becomes available~--- leverage that computing power to break today’s non-quantum-resistant encryption and gain access to such sensitive data. Given that some types of sensitive data need to remain confidential for many decades (e.g., state secrets), this is an issue that needs to be addressed sooner rather than later.   

The sooner the developers start working on upgrading their encryption schemes to the quantum-resistant ones, the sooner they will be able to mitigate this risk. Based on the timeline of the National Institute of Standards and Technology (NIST), we may expect to see a draft of the standard between 2022 and 2024~\cite{nist_timeline}. Nevertheless, it is still valuable to start upgrading cryptography-related code now, particularly doing so while adhering to crypto-agility practices (as recommended by NIST~\cite{chen2016report}), so that any algorithm chosen today can be changed in the future (if need be) without incurring a considerable cost.

While we need to wait two to four years to see the standard, we already have a good understanding of the classes of algorithms that will appear in this standard. In January 2019, NIST selected 26 candidates algorithms in Round 2 of the competition and will further trim this set in the subsequent rounds~\cite{nist_timeline}. 

Thus, software development organizations may start planning the migration now (e.g., by identifying vulnerable hardware and software components and  designing replacement schemes). In an enterprise environment with a large amount of hardware and software deployments, it may take many months (if not years) to identify all problematic  components and devise corrective actions. If an organization starts working on these tasks now, they may finish them by the time when the security standards will start to emerge.

\section{The impact of quantum computing on encryption algorithms}
It is essential to understand that quantum computing will affect encryption differently depending on the class of encryption algorithm. 

Below we first elaborate on the asymmetric encryption algorithms, which are used in many areas ranging from the Transport Layer Security (TLS) protocol (used to safeguard data passed between two systems on the Internet) to Pretty Good Privacy (PGP) software used to encrypt and decrypt a file and safely transfer it between computers.  

We then proceed to the symmetric encryption algorithms, such as Advanced Encryption Standard (AES), used to protect sensitive data. There exist numerous use-cases, ranging from encrypting a file archive (e.g., implemented in 7z and WinZip software) to encrypting computer’s disks (e.g, using Apple MacOS FileVault and Symantec Endpoint Encryption).

\subsection{Asymmetric encryption}

\begin{table*}[ht]
	\centering
	\caption{Effective security strength of key encryption algorithms as per~\cite{mavroeidis2018impact}}
	\label{tab:key}
	\resizebox{\textwidth}{!}{
		\begin{tabular}{lrrr}
			\toprule
			\multicolumn{1}{c}{Encryption algorithm} & \multicolumn{1}{c}{Key size (bits)} & \multicolumn{1}{c}{Effective security level on CCs (bits)} & \multicolumn{1}{c}{Effective security level on QCs (bits)} \\
			\midrule
			RSA 1024 & 1024 & 80 & 0 \\
			RSA 2048 & 2048 & 112 & 0 \\
			ECC 256 & 256 & 128 & 0 \\
			ECC 384 & 384 & 256 & 0 \\
			AES 128 & 128 & 128 & 64 \\
			AES 256 & 256 & 256 & 128 \\
			\bottomrule
		\end{tabular}
	}
\end{table*}

Asymmetric encryption algorithms, which are based on factoring large integers (e.g., Rivest–Shamir–Adleman~--- RSA), discrete logarithms (e.g., Elliptic Curve Cryptography~--- ECC, and Diffie-Hellman key exchange), or similar approaches (see~\cite{mavroeidis2018impact, schwabe2019} for review)  will need to be replaced by quantum-resistant alternatives~\cite{mavroeidis2018impact, schwabe2019}. Effective security strength, shown in Table~\ref{tab:key}, suggests that the strength of the RSA and ECC is somewhat weaker or comparable to AES on a CC, but is extremely weak on a QC. This is because Shor’s algorithm~\cite{shor1997} can perform integer factorization in polynomial time; so what requires thousands of years with classical computers would only take days/hours on a large-scale quantum computer. This, of course, assumes that a large-scale quantum computer with the required number of qubits exists, which is not the case right now. 

\subsection{Symmetric encryption}
Unlike asymmetric encryption algorithms, symmetric encryption algorithms do not face an existential threat: one needs to perform a brute-force attack to break it. However, on a CC generation of $n$ keys require $O(n)$ operations, while on a QC it can be done using $O\left(\sqrt{n}\right)$ operations, thanks to Grover’s algorithm. Thus, a large quantum computer running Grover’s algorithm could provide a quadratic improvement in brute-force attacks on symmetric encryption algorithms, such as AES. This translates into a need to double key size to support the same level of protection. For AES specifically, this means using 256-bit keys to maintain today’s 128-bit security strength\footnote{In other words, an $n$-bit AES cipher provides a security level of $n/2$  because $\sqrt{2^n} = 2^{n/2}$.}, as depicted in Table~\ref{tab:key}.

\section{The impact on security-critical systems}\label{sec:impact}
\subsection{Newly-built systems}\label{sec:new}
\subsubsection{Asymmetric encryption}
For new systems involving a security component, practitioners will have to replace modern asymmetric algorithms (e.g., RSA) with those that are based on algebraic operations, which QC cannot perform efficiently (in comparison with a CC). The field of post-quantum (also known as quantum-resistant) cryptography deals with such algorithms. Examples of the principles include but are not limited to lattice-based cryptography (e.g., used in NTRU~\cite{hoffstein1998ntru} and BLISS~\cite{ducas2013lattice} cryptographic schemes) and hash-based cryptography (e.g., using Merkle Hash Tree signature~\cite{merkle1987digital}). 

The efforts are underway to introduce cryptographic standards for the era of quantum advantage. As discussed in Section~\ref{sec:cta}, NIST is planning to come up with the standard between 2022 and 2024.

Thus, soon, a practitioner may be able to leverage a standardized algorithm, while designing new software. For now, when creating a new software product that is expected to have a lifespan long enough to be affected by the quantum advantage, we advise designing security component in such a way that the underlying cryptographic algorithm can be replaced with a different one (i.e., by using crypto-agility practices). 

\subsubsection{Symmetric encryption}
If the system requires a symmetric algorithm, then one can leverage a standard implementation of an existing algorithm (e.g., AES). In this case, the component has to be designed in such a way that it can accommodate the increase of the key length for a given algorithm. 

\subsection{Threats to the existing data}\label{sec:data}
While large-scale quantum computers might be several years away, someone with malicious intent could still capture sensitive encrypted data (e.g., by capturing encrypted network packets protected with TLS or by cloning an encrypted disk), then store that data somewhere in a data lake. When a large-scale quantum computer becomes available, this person can leverage QC power to break the asymmetric encryption used by TLS or brute-force access to the encrypted disk and recover the sensitive data. While not much can be done about the protocols involving asymmetric algorithms; for the symmetric ones, we can increase the length of the key right away to cumber the brute-force attack~\cite{Muppidi2018}. We can also encrypt archived data (e.g., stored on backup devices) with a quantum-resistant algorithm.

Another example is a blockchain platform using proof-of-work algorithms (e.g., based on SHA-256). The security of current blockchain platforms relies on a digital signature, which is based on either Elliptic Curve Digital Signature Algorithm~\cite{gupta2017blockchain} or RSA algorithms; both are vulnerable to QCs. Kiktenko et al.~\cite{kiktenko2018quantum} proposed a quantum-secured blockchain framework that utilizes Quantum Key Distribution techniques via an experimental fibre network (the cost of the network is not disclosed). An alternative is to introduce a quantum resistant asymmetric algorithm and recompute the Nonce for all legacy blocks using new algorithms, which may be expensive. A more efficient approach may be to switch to a proof-of-stake algorithm, which does not require solving computationally intensive mathematical puzzles, from a proof-of-work algorithm.

\subsection{Legacy systems}\label{sec:legacy}
If the legacy system is well-designed and actively maintained, then the solution is straightforward: one can replace an existing asymmetric algorithm with a new one (or increase a key size of an asymmetric one) while ensuring that the existing data can be migrated to a new format. However, it may require a downtime to re-encrypt existing data with a quantum-resistant algorithm, and re-encrypting data is typically disruptive until the new encrypted data are available again.

The software may run on antiquated hardware that does not have sufficient computing power to run QC-resistant algorithms. In this case, we may need to upgrade this obsolete equipment or virtualize the outdated runtime environment, so that it can be executed by a hypervisor on a more powerful computer. 

Often, altering existing (legacy) system to address the security concern may be challenging. Legacy systems frequently lack adequate information or support to be maintained or upgraded~\cite{khadka2014professionals}. The root causes of these issues are numerous. For example, developers of the system may be unavailable (e.g., because they left the company or retired), source code or documentation may be lost, or build platforms for the source code may be sunset. To make matters worse, the encryption-related code may be spread or cloned among multiple software components (due to bad design), making alterations even more challenging. These root causes make it extremely difficult and expensive to upgrade such a system to the newest security protocols, making the replacement the only feasible option.

\section{Quantum advantage and Y2K bug: parallels}

All of these challenges, conceptually, pose a striking similarity to the Y2K bug~\cite{britannica_y2k,Miranskyy2019}, which happened around the year 2000. The root cause of the Y2K bug was because older software represented four-digit years with only the last two digits, while the first two digits were fixed at 19. That is an increment by 1 of the year 1999 would result in the year 1900 rather than the year 2000. 

Obviously, the underlying root cause of the Y2K bug and the quantum advantage problems are different. Three critical differences between these two problems (in terms of nature, method and time) are as follows. 
\begin{enumerate}
	\item The Y2K bug was threatening the integrity and availability of data in computer systems. The quantum advantage threatens the confidentiality of data as well as the validity of data integrity and users' authentication. 
	\item In the Y2K case, we had to take timing-related code and replace it with a new one capable of representing years using four digits. In the quantum advantage case, we will have to take an encryption-related code and either replace the algorithm (asymmetric case) or increase key length (symmetric case).
	\item Finally, the failures associated with the Y2K-related defect were encountered after a particular date, namely, December 31, 1999.  In the case of quantum advantage, the exact encounter date is unknown. However, it is imminent, once powerful enough QCs are available. Moreover, the existing data are already at risk (as discussed in Section~\ref{sec:data}).
\end{enumerate}

Despite the differences, the conceptual solutions to the problems are similar. That is, either fix an existing code or make sure that the new code is resistant to the threat (be it the Y2K bug or quantum advantage). Thus, we may still leverage lessons learnt from fixing legacy systems to fix the Y2K bug. Let us review what we have learned from Y2K. Putnam and Myers~\cite{putnam1999year} divided the legacy systems into three groups for the Y2K bug. The same grouping strategy can be adopted for the quantum advantage problem: 1) work first on those that involve life and death, 2) work on those that are critical to the continued operation of your organization, and 3) work on those in which security is merely irritating, not costly. 

As the era of quantum computing is approaching, some actions can be taken now. Shimeall et al.~\cite{shimeall1999software} proposed several guidelines for the security concerns before Y2K, some of which are still applicable to our current situation: 1) existing systems must be examined and repaired, and 2) programmers and designers must be educated about the new security challenges brought by quantum computers. The second guideline applies to the creation of the new systems too.

Schultz~\cite{schultz1998managing} described five steps that a software engineer should follow to respond to the Y2K bug, which we can utilize in our challenge: 1) gain senior management’s acknowledgement of the potential impacts, 2) assess the problem and alternative solutions, 3) estimate the cost and gain approval of selected solutions, 4) execute and control the solutions as a partner with senior management, and 5) monitor solutions’ early results in production.

\section{Roadmap}
We propose a \textbf{7E} roadmap for software developers, summarizing steps needed to address encryption-related challenges associated with quantum advantage. Our 7E roadmap consists of the following seven steps:

\begin{enumerate}
    \item \textbf{E}ngage executives and senior management so that they can sponsor the initiative. It is important to get the acknowledgement from the decision makers in your company or organization. Moreover, executives and senior management can assess security concerns from a different perspective. Quantum computing is an emerging technology, and it may take time for executives to learn its growing powers and realize the side effect of these powers on the existing cryptography schemes. To educate the management, you can use formal presentations or reports and incorporate feedback from them later on. 

    \item \textbf{E}xamine existing products and their cybersecurity components to identify and locate the issues, review the document and/or programs and assess the problems. For legacy systems, there exist difficult scenarios, such as lack of documentation, source code, or build infrastructure (as discussed in Section~\ref{sec:legacy}). Identify existing data (if any) that may requires protection. 

    \item \textbf{E}volve: design a new software with crypto-agility (as per NIST recommendations~\cite{chen2016report}) in mind, so that quantum-resistant algorithm can be added to the software later on. Conceptually, this should be achieved by designing the systems in such a way that an existing encryption scheme can be easily replaced with a new one. Moreover, the systems should be able to recognize and `translate' multiple encryption schemes.

	For example, the encryption component can be designed to be plug-and-play, so that an existing encryption algorithm can be replaced (if it is discovered to be vulnerable) with a robust one; we will provide further details in Section~\ref{sec:agility}. This will save costs in the future, when the standards for quantum-resistant encryption algorithms are finalized and our software has to be updated with these algorithms.
	
	The challenges (identified in the previous steps and in Section~\ref{sec:impact}) may affect the easiness of design decisions. For example, if documentation is missing, one may first have to reverse engineer the architecture from the code by extracting caller-callee function pairs and identifying security-related ones.
	
	For cryptography that involves multiple business partners, achieving crypto-agility requires that all business partners update hardware and software promptly. Moreover, all the partners should disclose crypto-related information, e.g., security certificates and protocols, to each other. We will discuss it further in Section~\ref{sec:org}.

    \item \textbf{E}ducate the programmers and designers to make sure that everyone is `on the same page’ because (in most cases) the security-related component are coupled with the remaining software components. This implies that the whole development organization needs to be aware of the challenges associated with quantum advantage. Security professionals often may not have a quantum physics background, instead they might come from software engineering or computer science field. Thus, many of them may not have sufficient knowledge of quantum computing to perform cryptographic migration confidently. This step is complementary to Step~1, with the focus on training technical staff rather than the management. 

    \item \textbf{E}stimate the impact of potential problems and the cost of alternatives to prioritize the problems. The findings from Steps 2 and 3 should help to estimate the cost. Rate the cost of potential solutions in terms of human and time resources. Work first on the systems that handle critical data first. The definition of critical will vary from industry to industry. A representative example is a system handling personal data, such as financial transactions or health records. 
    
    Upgrade of the cryptographic schemes involving symmetric encryption will typically be cheaper than the asymmetric one. The former will usually require increasing the length of the key and increasing the storage space for the keys. The latter may also require replacing the code needed to do the encryption.
    
    \item \textbf{E}xecute the new cybersecurity policy. Select and adopt appropriate solutions based on requirements, budgets, and priorities. As discussed in Section~\ref{sec:impact}, practitioners can execute the new cybersecurity policy in different ways. For newly-built systems, post-quantum cryptography may be adopted (see Section~\ref{sec:new}). For legacy systems, the software and associated hardware may have to be altered (see Section~\ref{sec:legacy}). For existing data, an intermediate solution~--- e.g., re-encrypting the existing data with a quantum-resistant cryptographic algorithm~--- may be applied (see Section~\ref{sec:data}). 

    \item \textbf{E}ssay the new cybersecurity policy. Keep monitoring the performance and the robustness of your new cybersecurity policy in production to make sure that the challenges associated with quantum advantage were addressed; adjust the policy if needed. Experiences and lessons learnt from one project may also apply to another one. These lessons could serve as a building block to a general theory of making encryption migration agile and smooth.
\end{enumerate}

\section{Discussion on crypto-agility}\label{sec:discussion}

\subsection{Evolvability}\label{sec:agility}

As discussed above, organizations can start planning the transition to future quantum-resistant cryptographic standards right now by making sure that their software architecture is crypto-agile. One primary feature of crypto-agility is agile evolution. 

A good example of evolving cryptographic components is given by Sullivan~\cite{sullivan2009cryptographic}. Assuming that we leverage Object Oriented programming, and each algorithm is implemented in a class, we can define general interfaces in abstract classes and then implement the interfaces in concrete classes. Say,  an abstract class \textit{HashAlgorithm} defines two methods \textit{ComputeHash} and \textit{Create}. We can easily implement a quantum-resistant hash algorithm, which inherits from this abstract class. The \textit{Create} method implements the Factory design pattern, which will enable swapping concrete classes by altering a configuration file rather than the source code. With this setup, a designer  can now go over their codebase and make sure that their use of hashing functions follows a similar pattern. Once they are ready to transition to the quantum-resistant hash function, it can be done with a single parameter switch. 

\subsection{Within- and cross-organizational collaboration}\label{sec:org}

Actors (within an organization or across multiple organizations) may use different encryption algorithms to pass the information. The data at rest may be encrypted by multiple algorithms. In this case, one needs to include information about the original encryption algorithm with the data, similar to Microsoft Office Document Cryptography Structure~\cite{sullivan2009cryptographic}. 

A common problem that cyber-security professionals face is a lack of understanding of where encryption schemes are applied throughout the infrastructure. Enterprise environments often have an extensive infrastructure with legacy components, making them hard to maintain and upgrade. In this case, crypto-agility best practices include establishing clear and transparent communication policies among different systems (e.g., installing complete certificate chains and configuring servers with the current TLS protocols). 

Multi-organizational interactions and collaborations are paramount and that they can be enabled by institutionalizing open standards, preferably with open source reference implementations of those standards. Conceptually, this process would be akin to the creation of other open standards, where an open standards organization would be formed to drive the process. 

Another example of such an organization  is the Internet Engineering Task Force (IETF), which works on Internet standards, including TLS protocol. In fact, IETF has already started working on making TLS quantum resistant. There exist multiple open source implementations of the TLS, such as OpenSSL and GnuTLS. One can create a fork of the implementation and use it for their needs. For example, LibreSSL is a fork of OpenSSL for the BSD community, while BoringSSL is a fork of OpenSSL for Google projects. Once IETF updates TLS with the quantum-resistant extensions, the changes will be implemented in the reference libraries, which will then ``trickle-down'' to the respective forks and clones.

\section{Summary}
To summarize, the community of software practitioners needs to start preparing for the era of quantum advantage on three fronts: 1) developing new software with new encryption algorithms in mind following crypto-agility practices, 2) upgrading the legacy software with new encryption algorithms, or 3) migrating the business processes to more modern software if legacy software cannot be upgraded. The earlier we start --- the higher the chances that we will be able to address proactively security challenges that this era brings with it.

\section*{Acknowledgment}
We thank the anonymous reviewers for the insightful comments and suggestions!

\bibliographystyle{IEEEtran}
\bibliography{IEEEabrv,reference}

\end{document}